\documentclass[reprint, aps, pre, superscriptaddress, longbibliography]{revtex4-2}

\usepackage{graphicx}
\usepackage{physics}
\usepackage{amsmath}
\usepackage{setspace}
\usepackage[caption=false]{subfig}
\usepackage{multirow}
\usepackage{dsfont}
\usepackage{xcolor}
\usepackage{amssymb}
\usepackage{mathtools}
\usepackage{mathrsfs}
\usepackage{cancel}
\usepackage{esvect}
\usepackage{indentfirst}
\usepackage{dcolumn}
\usepackage[colorlinks=true, allcolors=blue]{hyperref}
\usepackage{hyperref}

\begin{document}

\title{Percolation and Criticality in Hyperuniform Networks}


\author{Yongyi Wang}
\affiliation {Department of Physics, Pennsylvania State University, University Park, PA 16802}

\author{Jaeuk Kim}
\affiliation{Department of Physics,  Princeton University, Princeton, New Jersey 08544, USA} 

\author{Yang Jiao}
\affiliation {Materials Science and Engineering, Arizona State University, Tempe, AZ 85287} \affiliation{Department of Physics, Arizona State University, Tempe, AZ 85287}

\author{Izabella Stuhl}
\affiliation {Department of Mathematics, Pennsylvania State University, University Park, PA 16802}

\author{Salvatore Torquato}
\affiliation {Department of Chemistry, Princeton University, Princeton, New Jersey 08544, USA} \affiliation{Department of Physics,  Princeton University, Princeton, New Jersey 08544, USA} \affiliation{Princeton Institute  of Materials, Princeton University, Princeton, New Jersey 08544, USA}
\affiliation{Program in Applied and Computational Mathematics, Princeton University, Princeton, New Jersey 08544, USA}

\author{R{\'e}ka Albert}
\email[correspondence sent to: ]{rza1@psu.edu}
\affiliation{Department of Physics, Pennsylvania State University, University Park, PA 16802} 

\date{\today}




\begin{abstract}
{Disordered hyperuniform many-particle systems exhibit an anomalous suppression of density fluctuations on macroscopic length scales relative to those of conventional disordered systems. Here we investigate percolation on Delaunay triangulation networks constructed from disordered stealthy hyperuniform (SHU) point configurations, a subclass of hyperuniform systems whose structure factor vanishes over a finite range of wavevectors near the origin and whose degree of stealthiness is controlled by $\chi$, with Poisson point configurations included for comparison. We consider a non-uniform bond percolation process in which bond occupation probabilities decrease with Euclidean edge length, and determine the percolation thresholds and critical exponents using finite-size scaling. We find that SHU networks exhibit lower percolation thresholds than Poisson networks. Notably, the percolation threshold of SHU networks decreases with the stealthiness parameter $\chi$, indicating that system-spanning connectivity emerges more readily as short-range order increases. All systems considered are consistent with the standard 2D percolation universality class, although the Poisson and low-$\chi$ SHU networks exhibit stronger corrections to scaling.} Our work extends previous studies on transport properties of SHU systems from continuum two-phase media to networks. These results open new avenues for optimizing the resilience of statistically homogeneous disordered networks.
\end{abstract}


\maketitle

\section{Introduction}
A \textit{hyperuniform} many-particle system, disordered or not, is one in which the number variance of particle centers within a spherical observation window of radius $R$ grows slower than the observation window volume $R^d$ in $d$-dimensional space for large $R$ \cite{To03,To18a}, i.e.,
\begin{equation}
\lim_{R\rightarrow\infty}\sigma_N^2(R)/v_1(R) = 0
\end{equation}
where $\sigma_N^2(R)\equiv\langle N(R)^2\rangle - \langle N(R)\rangle^2$ is the variance in the number of particle centers (brackets indicate the ensemble average) and $v_1(R) \sim R^d$ is the volume of a $d$-dimensional sphere of radius $R$. Equivalently, a system is hyperuniform if its structure factor $S(\mathbf{k})$ (defined in Sec.~\ref{sec:HU_def}) tends to zero as the wavenumber $k\equiv|\mathbf{k}|$ tends to zero, i.e.,
\begin{equation}\label{eqn:S(k)}
    \lim_{|\mathbf{k}|\rightarrow0}S(\mathbf{k})=0.
\end{equation}
The small-$|{\bf k}|$ scaling behavior of $S({\bf k}) \sim |{\bf k}|^\alpha$ determines the large-$R$
asymptotic behavior of $\sigma_N^2(R)$, based on which all hyperuniform
systems, disordered or not, can be categorized into three classes:
$\sigma_N^2(R) \sim R^{d-1}$ for $\alpha>1$ (class I); $\sigma_N^2(R)
\sim R^{d-1}\ln(R)$ for $\alpha=1$ (class II); and $\sigma_N^2(R)
\sim R^{d-\alpha}$ for $0<\alpha<1$ (class III) \cite{To18a}. Consequently, hyperuniform systems encompass all crystals and quasicrystals \cite{To03, To18a}. Hyperuniformity has also been identified in a wide spectrum of physical \cite{ref4, ref5, ref6, ref7, ref16, ref17, ref18, ref19, ref20,
ref21, ref22, ref23, salvalaglio2020hyperuniform, hexner2017noise, hexner2017enhanced, weijs2017mixing,
lei2019nonequilibrium, lei2019random, ref8, ref9, ref10, ref11,
ref12, ref13, ref14, ref15, ref24, ref25, sanchez2023disordered}, material \cite{ref28, ref29, ref30, Ge19, sakai2022quantum, Zh20, Ch21, PhysRevB.103.224102, Zh21, nanotube, zhang2023approach, chen2021multihyperuniform, chen2025anomalous, shi2025three, wang2025hyperuniform, zhong2025modeling} and biological \cite{ref26, ref27, ge2023hidden, liu2024universal, tang2024tunablehyper} systems. The unique combination of local disorder and complete suppression of global density fluctuations endows these systems with unexpected physical properties, including novel wave propagation characteristics \cite{ref31, ref32, ref33, scattering, granchi2022near, park2021hearing, klatt2022wave, tavakoli2022over, cheron2022wave, yu2021engineered, li2018biological}, thermal, electrical and diffusive transport properties \cite{ref34, torquato2021diffusion, maher2022characterization}, mechanical properties \cite{ref35, puig2022anisotropic} as well as optimal multifunctional characteristics \cite{ref36, kim2020multifunctional, torquato2022extraordinary}. 

Stealthy hyperuniform (SHU) systems, the focus of this work, are a special subset of class-I hyperuniform systems, which possess a zero structure factor for a range of wavevectors around the origin, i.e., $S({\bf k}) = 0$ for ${\bf k} \in \Omega$ (excluding the forward scattering). Thus, in contrast to standard hyperuniform systems that suppress density fluctuations only at the infinite wavelength, SHU systems suppress density fluctuations from intermediate to infinite wavelengths. This unique structural feature endows disordered SHU systems with highly efficient transport characteristics despite the absence of crystalline order \cite{ref34, torquato2021diffusion, ref36}. 

In particular, a previous study of continuum percolation in two-phase media found that SHU two-phase media exhibit a higher percolation volume fraction than Poisson (uncorrelated) systems, with the threshold depending systematically on the stealthiness parameter $\chi$ \cite{ref34}. Klatt \textit{et al.} also investigated void percolation in ordered and disordered sphere configurations through a mapping to weighted bond percolation on the corresponding Voronoi networks, and found that the void-percolation threshold decreases with increasing structural order \cite{Klatt_percolation}. 

Motivated by these results, we investigate whether networks constructed from SHU point configurations likewise exhibit distinct percolation behavior. We generate Delaunay triangulation networks from SHU and Poisson point configurations and study bond percolation on the resulting graphs. The Delaunay triangulation provides a parameter-free planar mapping from a point configuration to a spatial network \cite{Delaunay_triangulation}. Unlike continuum percolation, where connectivity is geometrically determined for a given point configuration \cite{yuan2026universal, torquato2013effect}, bond percolation introduces an additional source of stochasticity through probabilistic edge occupation. Such models are widely used to study resilience and robustness in complex networks, including infrastructure and biological systems \cite{Newman_perco, Internet_perco, RevModPhys.80.1275, PhysRevE.106.014304}.



We examine a distance-dependent bond percolation on Delaunay triangulation networks derived from SHU point configurations. The vertices of these networks correspond to the points, and the edges of the triangulation are weighted by the Euclidean distance between corresponding vertices. Since the geometric distance between points encodes crucial information about density fluctuations that is difficult to infer from purely topological network measures, such as the degree distribution \cite{HU_Eli, HU_Eli_2}, we introduce a parameterized bond percolation model (see Sec.~\ref{sec:percolation_model}, Eq.~(\ref{eqn:perco_model})) that captures Euclidean distances. In this model, low-weight edges (corresponding to nearby points) have a higher probability of being occupied, as observed in many real-life networks. This assumption allows the model to provide insights into cluster formation in disordered systems and an emergent order on a global scale, as shown in the subsequent sections.


We show that adding global order to a disordered system can alter its critical behavior by determining the percolation threshold and the set of critical exponents for SHU systems with various $\chi$ values and for the uncorrelated Poisson system. We find that under the parameterized bond percolation model, the percolation threshold decreases with increasing degree of hyperuniformity. Our results demonstrate that SHU networks can achieve global connectivity at a lower coupling scale than the uncorrelated random system. Moreover, we estimate the critical exponents and fractal dimension of all networks. We find that all systems considered are consistent with the standard 2D percolation universality class. Although the Poisson and low-$\chi$ SHU networks exhibit apparent deviations in critical exponents obtained from leading-order finite-size scaling, these deviations are consistent with stronger corrections to scaling.

These findings have multiple implications. Our results suggest that Delaunay triangulation networks derived from SHU point configurations are more resistant to distance-dependent edge removal than those derived from Poisson point configurations.  Furthermore, we show that the percolation threshold of SHU networks is sensitive to the short-range order of the underlying point configurations, with the percolation threshold decreasing as short-range order increases. This behavior is consistent with the near-optimal conductive transport properties of high-$\chi$ SHU networks observed in Ref.~\cite{ref36}. Lastly, the degree of stealthiness $\chi$ generates a family of SHU networks with tunable critical thresholds. This provides a controlled setting for studying how structural correlations influence nonuniversal percolation properties. 

The rest of the paper is organized as follows: In Sec. \ref{sec:HU_def}, we provide definitions of concepts associated with hyperuniformity and correlation functions. In Sec. \ref{sec:percolation_model} we introduce our parameterized bond percolation model. Section \ref{sec:Methods} presents our numerical methods for studying the percolation behaviors of the SHU and Poisson networks. In Sec. \ref{sec:results}, we present our numerical results, including estimates of the percolation threshold and the critical exponents. In Sec. \ref{sec:conclusion_and_discussion}, we present concluding remarks. 

\section{Definition of Hyperuniformity} \label{sec:HU_def}
A point configuration in $d$-dimensional Euclidean space $\mathbb{R}^d$ is completely characterized by an infinite set of $n$-point correlation functions $\rho_n(\mathbf{r}_1,\dots,\mathbf{r}_n)$, each of which is proportional to the probability of finding $n$ points (representing the centers of particles) at the positions $\mathbf{r}_1,\dots,\mathbf{r}_n$ \cite{torquato2002random}.
For statistically homogeneous systems, $\rho_1(\mathbf{r}_1)=\rho$, and $\rho_2(\mathbf{r}_1,\mathbf{r}_2)=\rho^2 g_2(\mathbf{r})$, where $\mathbf{r}=\mathbf{r}_1-\mathbf{r}_2$, and $g_2(\mathbf{r})$ is the pair correlation function.
If the system is also statistically isotropic, then $g_2(\mathbf{r})=g_2(r)$, where $r=|{\bf r}|$.
The ensemble-averaged structure factor $S(\mathbf{k})$ is defined as
\begin{equation}
    S(\mathbf{k})=1+\rho\Tilde{h}(\mathbf{k})
\end{equation}
where $\Tilde{h}(\mathbf{k})$ is the Fourier transform of the total correlation function $h(\mathbf{r})=g_2(\mathbf{r})-1$, and ${\bf k}$ is the wave vector.

For a single periodic point configuration with $N$ particles at positions $\mathbf{r}^N = (\mathbf{r}_1,\dots,\mathbf{r}_N)$ within a fundamental cell $F$ of a lattice $\Lambda$, the scattering intensity $\mathbb{S}(\mathbf{k})$ is given by
\begin{equation}\label{eq:Skcomp}
    \mathbb{S}(\mathbf{k}) = \frac{|\sum_{j=1}^N \textrm{exp}(-i\mathbf{k}\cdot\mathbf{r}_j)|^2}{N}.
\end{equation}
In the thermodynamic limit, the scattering intensity of an ensemble of $N$-particle configurations in $F$ is related to $S(\mathbf{k})$ by
\begin{equation}
    \lim_{N,V_F\rightarrow\infty}\langle \mathbb{S}(\mathbf{k})\rangle = (2\pi)^d \rho \delta(\mathbf{k}) + S(\mathbf{k}),
\end{equation}
where $\rho = N/V_F$, $V_F$ is the volume of the fundamental cell, and $\delta$ is the Dirac delta function \cite{To03}.
For finite-$N$ simulations under periodic boundary conditions, Eq.~(\ref{eq:Skcomp}) is used to compute $S(\mathbf{k})$ directly by averaging over configurations.

Consider systems characterized by a structure factor with a radial power law in the vicinity of the origin,
\begin{equation}
    S(\mathbf{k})\sim|\mathbf{k}|^{\alpha}\;\textrm{for}\;|\mathbf{k}|\rightarrow0.
\end{equation}
The exponent $\alpha$ is referred to as the hyperuniformity exponent. For {\it hyperuniform} systems, $\alpha > 0$. A (standard) {\it nonhyperuniform} system has $\alpha = 0$, i.e., S({\bf k}) approaches a non-zero constant in the zero-wavenumber limit. An {\it antihyperuniform} system is one possessing a diverging S({\bf k}) in the zero-wavenumber limit, i.e., with $\alpha < 0$.

For hyperuniform systems, $\alpha$ determines large-$R$ scaling behaviors of the number variance \cite{To03, To18a}, according to which all hyperuniform systems can be categorized into three different classes:
\begin{equation}\label{eq:classes}
    \sigma^2_N(R)\sim
    \begin{cases}
    R^{d-1}&\alpha > 1, \textrm{class I}\\
    R^{d-1}\textrm{ln}(R)&\alpha = 1, \textrm{class II}\\
    R^{d-\alpha}&\alpha < 1, \textrm{class III}.
    \end{cases}
\end{equation}
Classes I and III are the strongest and weakest forms of hyperuniformity, respectively. Class I systems include all crystal structures \cite{To03}, many quasicrystal structures \cite{Og17}, and certain exotic disordered systems \cite{Za09, Ch18a}. Examples of Class II systems include some quasicrystal structures \cite{Og17}, perfect glasses \cite{zhang2017classical}, and maximally random jammed packings \cite{ref4, ref5, ref6, Za11c, Za11d}. Examples of Class III systems include classical disordered ground states \cite{ref8}, random organization models \cite{ref20}, perfect glasses \cite{zhang2017classical}, and perturbed lattices \cite{Ki18}; see Ref. \cite{To18a} for a more comprehensive list of systems that fall into the three hyperuniformity classes.


SHU systems are a special subset of class-I hyperuniform systems possessing a zero structure factor for a range of wavevectors around the origin, i.e.,
\begin{equation}
S({\bf k}) = 0, \quad\text{for}\quad {\bf k} \in \Omega,
\label{eq_stealthy}
\end{equation}
excluding the forward scattering. SHU systems include all crystals, most quasicrystals and certain special disordered systems \cite{To18a}. The degree of stealthiness and short-range order in the system is determined by a tuning parameter $\chi$ measuring the fraction of the independently constrained degrees of freedom $M$ within the exclusion region $\Omega$ (i.e., half the number of ${\bf k}$ points in $\Omega$) \cite{ref9, ref14, ref15, Zh17}, i.e.,
\begin{equation}
    \chi = \frac{M}{(N-1)d}
    \label{eq_chi_ratio}
\end{equation}
where $N$ is the total number of points in the systems. In this work, we focus on the disordered regime, i.e., $0 <\chi <1/2$, in which the systems possess no long-range order. We note that $d$ degrees of freedom associated with the trivial overall translation of the entire system are subtracted in Eq.~(\ref{eq_chi_ratio}). SHU systems distinguish themselves from standard hyperuniform systems in that they completely suppress density fluctuations at the intermediate to infinite wavelengths. In contrast, standard hyperuniform systems only completely suppress infinite-wavelength density fluctuations. 

\section{Parameterized bond percolation model} \label{sec:percolation_model}
 Consider a Delaunay triangulation graph $G$ corresponding to a point configuration, in which each edge is weighted by the Euclidean distance between the two connected vertices.  We propose the following bond percolation model: in $G$, each edge $e_{ij}\in E(G)$ is retained independently with probability
\begin{equation}\label{eqn:perco_model}
    p_{ij}=\max\left(0, 1-\frac{d_{ij}}{z}\right)
\end{equation}
where $d_{ij}$ is the Euclidean distance between two connected vertices $i$ and $j$, and $z$ is the tuning parameter that plays a role analogous to temperature in the Ising model or the occupation probability $p$ in Bernoulli percolation. In other words, each edge has a probability $1-p_{ij}=\min\left(1, \frac{d_{ij}}{z}\right)$ of being absent. This model reflects the physical assumption that longer edges are less likely to be present. Indeed, interactions limited by distance are observed in many real-life networks, including epidemic contact networks and neuronal signaling \cite{epi_net1, epi_net2, neuro_net1, brain_net1}. 

Effectively, we map distances between vertices onto an interaction network, where edge presence is governed by the distance-dependent probability $p_{ij}$. For $z \geq \max(d_{ij},d_{mn})$, the edge-absence probabilities satisfy $(1-p_{ij})/(1-p_{mn}) = d_{ij}/d_{mn}$, such that relative edge lengths are directly reflected in the bond occupation probabilities. The parameter $z$ sets the global scale of the interaction strength. In the limit $z \rightarrow 0$, there are no edges in the network. In the limit $z \rightarrow \infty$, all edges of the Delaunay triangulation are present. There is a phase transition at some finite $z_c$, above which a percolating cluster is observed.


\section{Methods} \label{sec:Methods}

We determine the percolation threshold $z_c$ and the corresponding set of critical exponents for the SHU networks with various $\chi$ values as well as for the Poisson networks using Monte Carlo simulations and finite-size scaling, as described below. 

\subsection{Generation of hyperuniform networks} 
Disordered SHU point configurations investigated in this work are generated using the so-called ``collective coordinates'' approach \cite{ref11, ref14, ref15}. Starting from a random initial configuration, the positions of the points are randomly perturbed to minimize the difference between the system's structure factor and the targeted structure factor within the exclusion region ${\bf k} \in \Omega$, via stochastic optimization. In this work, we include SHU systems with $\chi=0.20, 0.40$ and 0.49. 

To perform finite-size scaling, as described in Sec.~\ref{sec:fss_method}, it is necessary to generate systems of different linear sizes. For a linear size $L$, $N=L^2$ points are generated in an $L\times L$ box. Due to the stochastic nature of the point configurations, we prepare $N_{real}=20$ disordered realizations for each linear size to compute ensemble-averaged observables. To balance the computational cost of system generation with the need for high statistics in the percolation calculations, we study linear sizes in the range $20\leq L \leq200$.   

\begin{figure*}[t]
\centering

\subfloat[Poisson\label{fig:PC_Poisson}]{
    \includegraphics[width=0.235\textwidth]{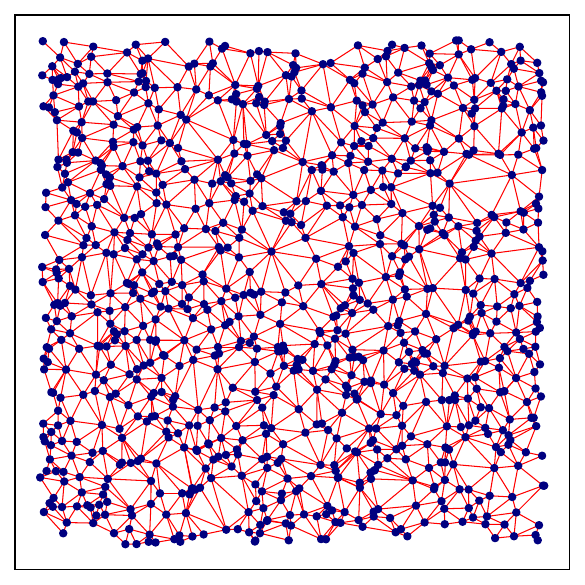}
}%
\hfill
\subfloat[SHU $\chi=0.20$\label{fig:PC_S20}]{
    \includegraphics[width=0.235\textwidth]{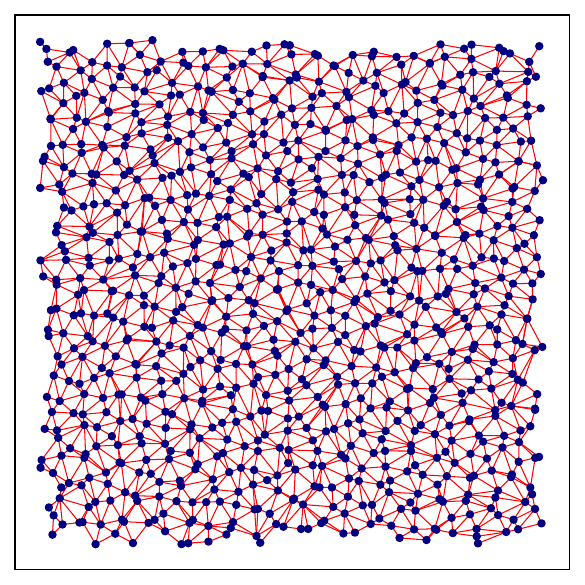}
}%
\hfill
\subfloat[SHU $\chi=0.40$\label{fig:PC_S40}]{
    \includegraphics[width=0.235\textwidth]{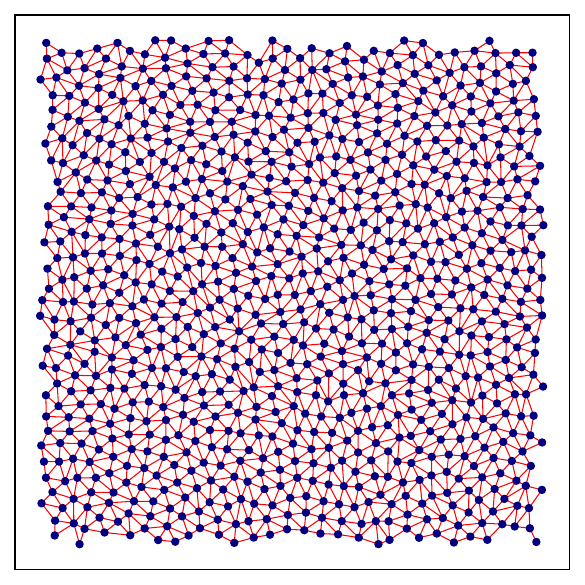}
}%
\hfill
\subfloat[SHU $\chi=0.49$\label{fig:PC_S49}]{
    \includegraphics[width=0.235\textwidth]{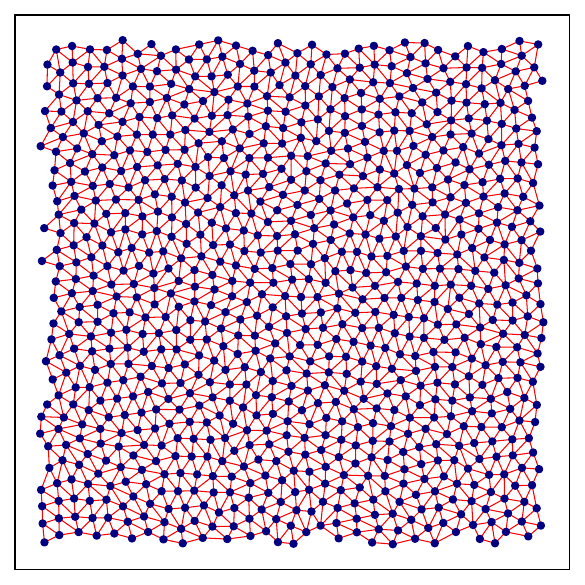}
}

\caption{
Delaunay triangulations of Poisson and SHU
point configurations. The vertices are shown as dark blue points,
and the edges are shown in red. Only a section of each periodic
network with toroidal topology is displayed.
}
\label{fig:all_PCs}

\end{figure*}

Once the SHU point configurations are obtained, we construct Delaunay triangulation networks with periodic boundary conditions, following Ref.~\cite{Delaunay_triangulation, HU_Eli_2}. In a Delaunay triangulation of a point configuration, the points become vertices of the graph, and edges are drawn between neighboring points such that the resulting graph is planar, triangular, and the circumcircles of each triangle do not contain any other points. Figure \ref{fig:all_PCs} shows realizations of the Delaunay triangulation of the three SHU systems, as well as that of an uncorrelated Poisson point configuration.

     

\subsection{Finite-size scaling}\label{sec:fss_method}
In finite systems, the growing correlation length near criticality is limited by the system size $L$. Finite-size scaling exploits the resulting size dependence to determine the thermodynamic-limit critical behavior \cite{St94}. An observable $Q(z)\sim|z-z_c|^{-q}$ in the thermodynamic limit scales as 
\begin{equation}
    Q_L(z)= L^{\frac{q}{\nu}}F\left(x\right)
\end{equation}
when the system is finite, where $F(\cdot)$ is some scaling function and $x=(z-z_c)L^{\frac{1}{\nu}}$. In particular, the probability that there exists a percolating cluster, $R_L(z)$, scales as 
\begin{equation}\label{eqn:R_L}
    R_{L}(z) = \Phi\left(x\right)
\end{equation}
which becomes a step function centered at $z_c$ as $L\rightarrow\infty$. $R_L(z)$ plays a central role in determining the percolation threshold $z_c$ and the exponent $\nu$. Knowing the value of the exponent $\nu$ allows the calculation of other critical exponents $q$ since $Q_L(z_c)\propto L^{\frac{q}{\nu}}$. 

First, $\nu$ can be calculated by noting that if one defines two constants $R_1$ and $R_2$ such that $R_L(z_1)=R_1$ and $R_L(z_2)=R_2$, Eq.~(\ref{eqn:R_L}) suggests that the width $\Delta z = z_1-z_2$ scales as 
\begin{equation}\label{eqn:transition_width_scaling}
    \Delta z \propto L^{-\frac{1}{\nu}}
\end{equation}
The scaling relation is independent of the choice of $R_1$ and $R_2$, which only contributes to the nonuniversal prefactor. Intuitively, this scaling relation can also be understood by noting that the correlation length $\xi\sim|z-z_c|^{-\nu}$ becomes comparable to $L$ at the percolation threshold, which implies $|z-z_c|\sim L^{-\frac{1}{\nu}}$. Therefore, the critical exponent $\nu$ can be evaluated with Eq.~(\ref{eqn:transition_width_scaling}) by calculating $\Delta z$ for different $L$. 

Second, $z_c$ is determined by making the observation that 
\begin{equation}\label{eqn:RL_slope}
    \frac{dR_L(z)}{dz} = L^{\frac{1}{\nu}}\Phi'(x)
\end{equation}
We define a finite-size pseudocritical point $z_c(L)$ where the slope $dR_L(z)/dz$ is maximal. According to Eq.~(\ref{eqn:RL_slope}), this occurs at $x^*=\left(z_c(L) - z_c\right)L^{\frac{1}{\nu}}$ where $x^*$ is a constant, and therefore, 
\begin{equation}
    z_c(L) = z_c + x^{*}L^{-\frac{1}{\nu}}
\end{equation}
Along with Eq.~(\ref{eqn:transition_width_scaling}), this yields
\begin{equation}\label{eqn:zc} 
    z_c(L)-z_c\propto \Delta z
\end{equation}
Hence, $z_c$ is determined by plotting $z_c(L)$ against $\Delta z$ and extrapolating to $\Delta z=0$. We compute $R_L(z)$ with the Newman-Ziff algorithm, as explained in the following section. 

The remaining critical exponents are determined from cluster observables
evaluated at the percolation threshold $z_c$. The fractal dimension $d_f$
characterizes how the mass $M(L)$ of the percolating cluster at criticality
scales with the linear size $L$,
\begin{equation}
    M(L) \sim L^{d_f}.
\end{equation}
The exponent ratio $\beta/\nu$ is obtained from the finite-size scaling of the fraction of vertices in the largest cluster,
\begin{equation}
    P_{\text{max}}(z_c) \sim L^{-\beta/\nu}.
\end{equation}
The critical exponent $\gamma$ characterizes the divergence of the mean finite-cluster size, excluding the percolating cluster, near $z_c$. At criticality, its finite-size scaling is
\begin{equation}
    S_L(z_c) \sim L^{\gamma/\nu}.
\end{equation}
Finally, scaling theory predicts the cluster-size distribution
\begin{equation}
    n_s(z)=q_0 s^{-\tau}
    f\left[q_1(z-z_c)s^\sigma\right],
\end{equation}
so that at the percolation threshold,
\begin{equation}
    n_s(z_c)\sim s^{-\tau}.
\end{equation}
In particular, we use the cluster distribution of systems with $L=200$ at criticality to estimate the exponent $\tau$. We introduce cutoffs $s_{\text{min}}=50$ and $s_{\text{max}}=5000$ to exclude the small-$s$ discreteness effects and the large-$s$ finite-size cutoff.

We sample the clusters at $z=z_c$ to determine these critical exponents. For each configuration, we calculate the cluster-size distribution $n_s$ and the percolating-cluster mass $M(L)$, from which the corresponding finite-size observables are obtained. The simulation is run $10^6$ times for $L\leq40$ and $10^5$ times for $L>40$ for each of the 20 disordered realizations of every system.

\subsection{Simulation with Newman-Ziff algorithm} 
The Newman-Ziff algorithm is a highly efficient Monte Carlo algorithm first proposed to determine the percolation threshold of Bernoulli percolation on lattices using the Union-Find data structure \cite{NZ_alg}. The standard Newman-Ziff algorithm samples configurations with a fixed number of occupied bonds (microcanonical ensemble) and recovers canonical observables through a binomial convolution \cite{NZ_alg}. In our model, however, the occupation probabilities depend on the edge lengths, so configurations with the same number of occupied bonds are not equally probable. We therefore propose the following algorithm that samples the canonical ensemble directly using edge-dependent activation thresholds.

For each run, we generate an independent random number $u_{ij} \sim \mathcal{U}(0, 1)$ for each edge $e_{ij}$, which is then assigned a corresponding threshold $z_{ij} = \frac{d_{ij}}{1-u_{ij}}$. At a given $z$, an edge $e_{ij}$ is present if $z_{ij}\leq z$. Since
\begin{equation}\label{eqn:sampling}
     p_{ij }= \mathbb{P}\left(u_{ij}\leq 1-\frac{d_{ij}}{z}\right) = \mathbb{P}\left(z_{ij}\leq z\right)
\end{equation}
this construction gives the desired occupation probability for every edge. Thus, at any given $z$, this algorithm directly samples the canonical ensemble of our percolation model, and no convolution is required.

We sort the edges in the ascending order of $z_{ij}$. Under periodic boundary conditions, we define a percolating cluster as one that wraps around the torus. This choice reduces boundary effects and improves the convergence of the percolation threshold \cite{NZ_alg}. We add edges sequentially in ascending order of $z_{ij}$ and check whether wrapping has occurred at each step. 

First of all, wrapping requires the formation of a non-contractible loop on the torus, which only occurs when a newly added edge connects two vertices that already belong to the same cluster. To further check whether the newly added edge results in a wrapping, we build upon the method used in continuum percolation \cite{Continuum_perco}. Whenever we use Union-Find to identify the root of a vertex $k$, we calculate the displacement $\vec{d}_{k}$ from $k$ to its root under periodic boundary conditions. When an edge $e_{ij}$ is added and $i$ and $j$ already share the same root, we calculate the loop displacement $\vec{D}_{loop}$ from $i$ to the root, the root to $j$, and $j$ to $i$:
\begin{equation}
    \vec{D}_{loop}=\vec{d}_i-\vec{d}_j+\Delta\vec{r}_{ij}
\end{equation}
where $\Delta\vec{r}_{ij}$ is the displacement from $j$ to $i$ under periodic boundary condition. This gives the winding number of the cluster to which $i$ and $j$ belong in both the $x$ and $y$ directions: 
\begin{equation}
    n_m =\frac{D_{loop}^m}{L},\quad m=x,y
\end{equation}
$n_m$ is a non-zero integer if wrapping in the $m$ direction is induced by the edge $e_{ij}$. Otherwise, $\vec{D}_{loop}=0$ and $n_m=0$. We study both horizontal and vertical wrapping, and define \cite{YSH}
\begin{equation}
    R_L(z)=\frac{1}{2}\left[R_{L,x}(z)+R_{L,y}(z)\right]
\end{equation}

We define $z_{\text{wrap},m}$ to be the wrapping threshold in the $m=x,y$ direction of the finite-sized system. For run $k$, $z^k_{\text{wrap},m}$ is the smallest $z_{ij}$ when wrapping first occurs in a given direction, and the contribution of that run to the wrapping probability is therefore
\begin{equation}
    R^k_{L,m}(z)=\textbf{1}\left(z^k_{\text{wrap},m}\leq z\right)
\end{equation}
where $\textbf{1}(\cdot)$ is an indicator function that equals 1 when the condition is true and is 0 otherwise. We record $z^k_{\text{wrap},m}$ for each run $k$. For $N$ runs, the wrapping probability is estimated as
\begin{equation}
    R_{L,m}(z) =\frac{1}{N}\sum_{k=1}^{N}\textbf{1}\left(z^k_{\text{wrap},m}\leq z\right)
\end{equation}
which converges to $R_{L,m}(z)=\mathbb{P}\left(z_{\text{wrap},m}\leq z\right)$.
Since $R_{L,m}(z)$ is the cumulative distribution function of $z_{\text{wrap},m}$, its derivative is the probability density of the wrapping threshold,
\begin{equation}\label{eqn:dR/dz and f}
    \frac{dR_{L,m}(z)}{dz}=f_{\text{wrap},m}(z),
\end{equation}
where $f_{\text{wrap},m}(z)$ denotes the probability density of $z_{\text{wrap},m}$. Therefore,
\begin{equation}
    \frac{dR_L(z)}{dz}
    =
    \frac{1}{2}
    \left[
    f_{\text{wrap},x}(z)+f_{\text{wrap},y}(z)
    \right]
\end{equation}
We estimate $dR_L(z)/dz$ directly from histograms of the recorded wrapping thresholds. 

For each disordered realization, we run the simulation $N=10^6$ times for $L\leq40$ and $N=10^5$ times for $L>40$. We then compute $\langle R_{L} (z)\rangle$ averaged over disordered realizations. 

\subsection{Data collapse optimization}\label{subsec:opt_refinement}
For a phase transition model that obeys finite-size scaling [see Eq.~(\ref{eqn:R_L})], $\langle R_{L} (z)\rangle$ for different $L$ are expected to collapse onto a single scaling function $\Phi(x)$ when plotted against $x=(z-z_c)L^{\frac{1}{\nu}}$ with the true $z_c$ and $\nu$ of the system. Data collapse serves as a consistency check of the finite-size scaling ansatz and of the estimated $z_c$ and $\nu$. 

We refine the initial estimate of $z_c$ from Eq.~(\ref{eqn:zc}) via a one-dimensional optimization of a collapse cost. We define a grid of $\{x_i=(z_i-z_c)L^{\frac{1}{\nu}}\}$ values in the critical region, from which $\langle R_{L} (x_i)\rangle$ could be calculated for all $L$. The collapse error is then defined to be the summed variance of $\langle R_{L} (x_i)\rangle$ across different linear sizes, and is minimized with respect to $z_c$ in the vicinity of the previously estimated value. We keep $\nu$ fixed to its independently determined value from the transition-width scaling in Eq.~(\ref{eqn:transition_width_scaling}), which is well constrained numerically as discussed in Sec.~\ref{sec:results}. We estimate uncertainties via bootstrap resampling of realizations, and report the mean and standard deviation of the bootstrap $z_c$ estimates. The refined $z_c$ values yield a stronger collapse onto a single universal scaling function, as shown in Sec.~\ref{subsec:zc}.

\section{Results} \label{sec:results}
We use the Newman-Ziff algorithm as described in Sec.~\ref{sec:Methods} to first determine the average wrapping probability $\langle R_{L} (z)\rangle$, from which we estimate the percolation threshold $z_c$. We then calculate the critical exponents and exponent ratios $\nu$, $\beta/\nu$, $\gamma/\nu$, and $\tau$, as well as the fractal dimension $d_f$, to characterize the universality class of the systems. The data that support the findings of this article are openly available \cite{wang_2026_19198330}. 

\subsection{The percolation threshold}\label{subsec:zc}
\begin{figure*}
\centering

\subfloat[Probability density $f_\text{wrap}(z)$\label{fig: dR/dz}]{
\includegraphics[width=0.48\textwidth]{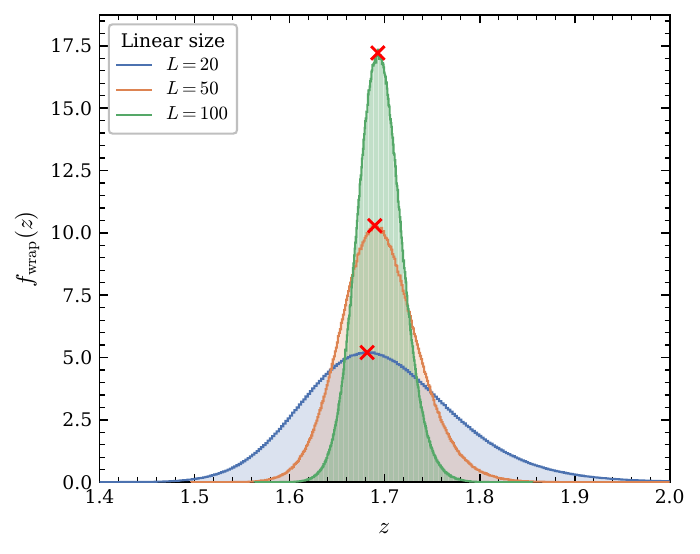}
}
\hfill
\subfloat[Finite-size scaling of $z_c(L)$\label{fig: crit pts}]{
\includegraphics[width=0.48\textwidth]{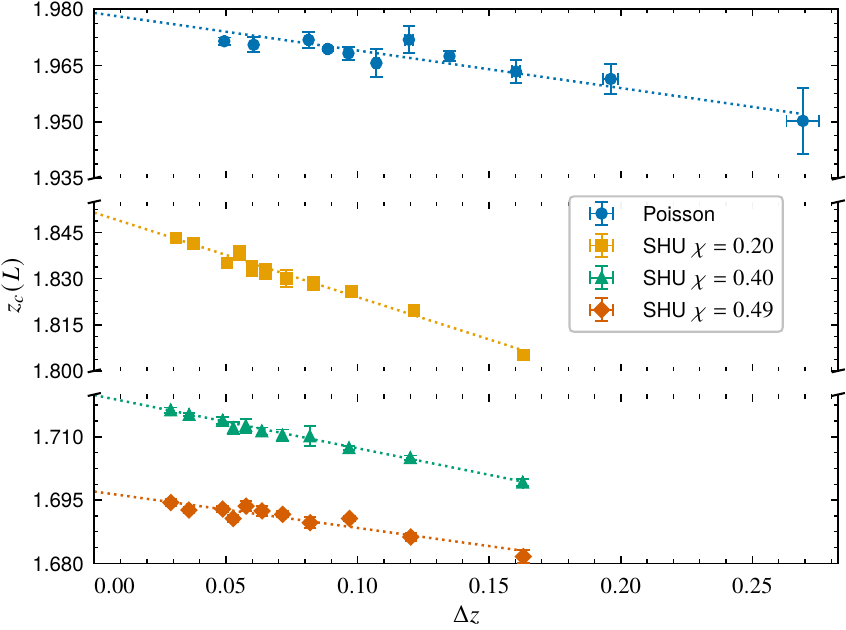}
}
\caption{Probability density of the wrapping thresholds for the SHU network with $\chi=0.49$ at various linear sizes in (a). Red crosses indicate the pseudocritical points $z_c(L)$, obtained from a quadratic fit to the histogram near its maximum. $f_\text{wrap}(z)$ gives the first derivative of the wrapping probability [see Eq.~\ref{eqn:dR/dz and f}]. (b) shows extrapolation of the pseudocritical point $z_c(L)$ to the thermodynamic limit ($\Delta z\rightarrow 0$) for all systems. The more ordered systems are found to have a lower percolation threshold. The uncertainties in $\Delta z$ reflect the cross-realization fluctuations of the crossing points $z_1$ and $z_2$. The uncertainties in $z_c(L)$ and $\Delta z$ decrease with the linear size $L$. }
\label{fig: dR/dz and fss}
\end{figure*}

The percolation threshold is estimated by extrapolating the pseudocritical point $z_c(L)$ as a function of the transition width $\Delta z$ [see Eq.~\ref{eqn:zc}]. First, the pseudocritical point $z_c(L)$ is computed by locating the peak of the histogram $f_{\text{wrap}}(z)$. To minimize the impact of noise, we perform a quadratic fit in the vicinity of the absolute maximum to find $z_c(L)$. Figure~\ref{fig: dR/dz} presents the histograms for various linear sizes, along with the corresponding pseudocritical points $z_c(L)$. 

We then determine the width of the transition $\Delta z$, as described in Sec.~\ref{sec:fss_method}. In theory, the calculations of $z_c$ and $\nu$ are independent of the choice of $R_1$ and $R_2$ that define $\Delta z$. Numerically, we find that choosing $R_1=0.85$ and $R_2=0.15$ leads to smaller uncertainties in $\Delta z$, while the transition width is still defined within the critical region. Nonetheless, the values of $z_c$ and $\nu$ obtained using different choices of $R_1$ and $R_2$ remain consistent within the error bars. Figure~\ref{fig: crit pts} shows the finite-size scaling of $z_c(L)$ across all four systems, highlighting a clear hierarchy: the more ordered the system, the lower its percolation threshold. 

The critical exponent $\nu$ is obtained from the slope of $\ln \Delta z$ versus $\ln L$, as given by Eq.~(\ref{eqn:transition_width_scaling}). The resulting estimates for all four systems are summarized in Table~\ref{table:crit_exp}. The estimates are stable under different choices of $R_1$ and $R_2$, providing well-constrained values of $\nu$ for the subsequent data-collapse analysis. 

\begin{figure}
\centering

\subfloat[Before optimization\label{fig:collapse_before}]{
\includegraphics[width=0.48\textwidth]{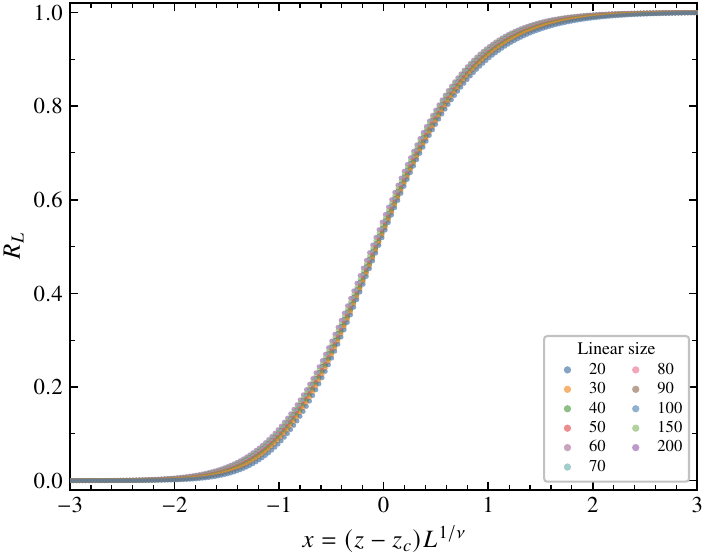}
}
\hfill
\subfloat[After optimization\label{fig:collapse_after}]{
\includegraphics[width=0.48\textwidth]{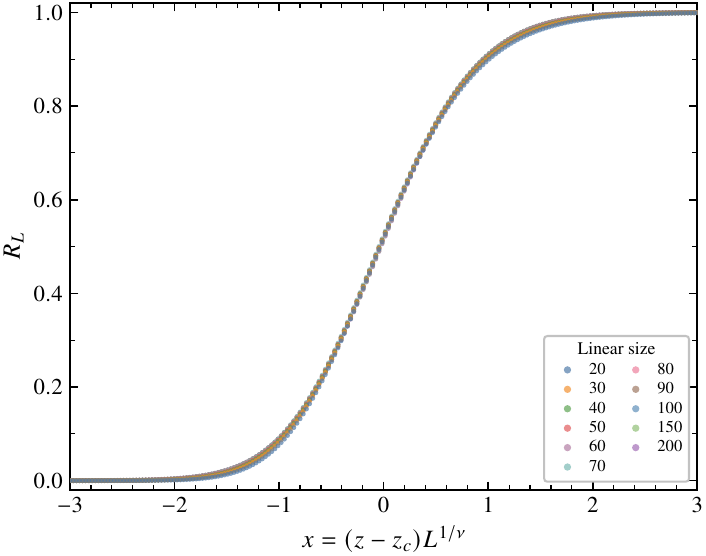}
}

\caption{The data collapse of SHU systems with $\chi=0.49$ is shown (a) before and (b) after applying the optimization algorithm described in Sec.~\ref{subsec:opt_refinement}. In (a), the collapse exhibits a small but visible misalignment, reflecting a deviation from the true $z_c$. After optimization (b), the collapse improves significantly, producing a more accurate estimate of $z_c$.}
\label{fig:collapse_S49}

\end{figure}

We assess the accuracy of the estimates of $z_c$ and $\nu$ by performing data collapse, i.e., checking whether the average wrapping probabilities $\langle R_L(z)\rangle$ for all linear sizes $L$ collapse onto a single universal scaling function. Figure~\ref{fig:collapse_S49} shows the data collapse of the SHU systems with $\chi=0.49$. As seen in Fig.~\ref{fig:collapse_S49}, the estimated $z_c$ and $\nu$ yield a reasonable collapse, but with a small misalignment. This is because the data collapse is very sensitive to $z_c$, whose estimation suffered from the noise, especially at small $L$, due to uncertainties in both $\Delta z$ and $z_c(L)$ [see Fig.~\ref{fig: crit pts}]. To improve the estimation of $z_c$, we employ the optimization algorithm detailed in Sec.~\ref{subsec:opt_refinement}. After optimization, the wrapping probabilities $\langle R_L(z)\rangle$ exhibit a stronger collapse onto a single scaling function as shown in Fig.~\ref{fig:collapse_after}. 

Table \ref{table:zc_and_nu} summarizes the estimated $z_c$ for all systems. We find that the SHU networks exhibit a lower percolation threshold $z_c$ than that of an uncorrelated random network, $z_c^{\mathrm{SHU}} < z_c^{\mathrm{Poi}}$. This means that SHU networks can achieve global connectivity at a lower coupling scale than the uncorrelated random network. We propose that this hierarchy of percolation thresholds arises from the suppression of density fluctuations in the underlying SHU point configurations. 

\begin{figure}[h]
\centering

\subfloat[Poisson]{
\includegraphics[width=0.23\textwidth]{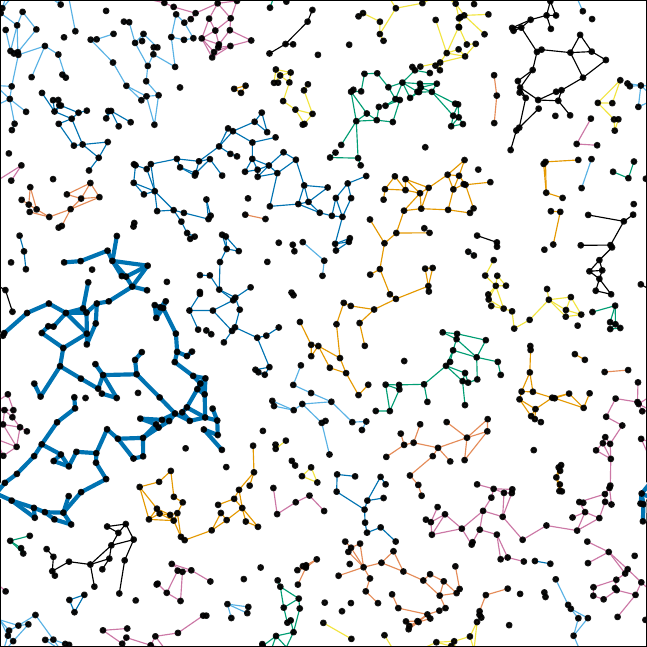}
}
\subfloat[SHU $\chi=0.40$]{
\includegraphics[width=0.23\textwidth]{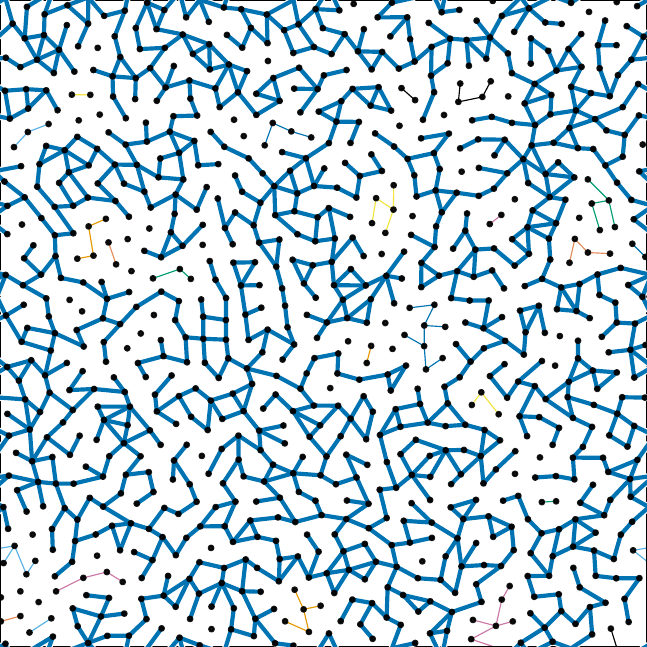}
}

\caption{Configurations of (a) Poisson and (b) SHU ($\chi=0.40$) systems at $z^{*}=1.8$, which lies between the percolation thresholds of the two systems under the percolation model given by Eq.~\ref{eqn:perco_model}. The system size is $30\times30$ for both systems, with periodic boundary conditions. Disjoint clusters are denoted by different colors. The largest cluster is highlighted with thick blue bonds. The Poisson configuration exhibits pronounced microclustering, which is strongly suppressed in the hyperuniform case.}
\label{fig:configs}

\end{figure}

Figure~\ref{fig:configs} supports this interpretation by showing configurations for Poisson and SHU ($\chi=0.40$) systems at an intermediate value \(z^\star=1.8\) such that $z_c^{\mathrm{SHU}} <z^{*}< z_c^{\mathrm{Poi}}$. Due to the distance dependence of the bond occupation probability, nearby sites are connected first as the coupling scale $z$ increases. For the Poisson system, this mechanism leads to highly localized clusters surrounded by sizable voids, and to achieve global connectivity, a higher coupling scale is required to bridge the gaps between local clusters. Conversely, hyperuniformity suppresses such microclustering, which shortens the required bridge lengths and yields a lower percolation threshold. In particular, it has been shown that disordered stealthy configurations across the first three space dimensions possess  ``bounded holes'' in the infinite-system-size limit, in contrast to Poisson systems which can possess arbitrarily large holes \cite{Zh17}. Similarly, previous studies of continuum percolation found that SHU systems percolate at a smaller critical particle radius than Poisson systems, consistent with the suppression of large holes in SHU configurations \cite{ref34}. Our results show that this mechanism extends from continuum media to SHU networks.


\begin{table}[h]
\caption{Percolation thresholds $z_c$ of the Poisson and the SHU networks with different $\chi$ values.}
\centering
\renewcommand{\arraystretch}{1.2}
\setlength{\tabcolsep}{14pt}   
\begin{tabular}{cc}
\hline
System & $z_c$ \\
\hline
Poisson & $1.97432(78)$ \\
SHU $\chi=0.20$ & $1.85012(118)$ \\
SHU $\chi=0.40$ & $1.71752(19)$ \\
SHU $\chi=0.49$ & $1.69540(9)$ \\
\hline
\end{tabular}
\label{table:zc_and_nu}
\end{table}

Interestingly, the percolation threshold of SHU networks could be controlled by varying the stealthiness parameter $\chi$: the higher the $\chi$ value, the lower the percolation threshold. This behavior arises because $\chi$ directly controls the degree of short-range order in the system, and microclustering is highly suppressed when all edges have comparable occupation probability, as it is the case for systems with pronounced local order. The non-universal percolation threshold $z_c$ is sensitive to such microscopic structural details. Therefore, stealthy hyperuniformity enables controlled manipulation of network criticality through structural correlations, offering new opportunities for optimizing connectivity in disordered systems.

\subsection{Critical exponents}\label{sec:universality_class}
To characterize the universality class of the percolation transition, we obtain effective estimates of the critical exponents $\nu$, $\beta/\nu$, $\gamma/\nu$, and $\tau$, as well as the fractal dimension $d_f$, using the leading-order scaling forms described in Sec.~\ref{sec:fss_method}. The resulting effective estimates for all systems are summarized in Table~\ref{table:crit_exp}, together with the exact values for standard 2D lattice percolation.

\begin{table*}[t!]
\caption{Effective critical exponents and fractal dimension of the Poisson and the SHU networks with various $\chi$.}
\centering
\setlength{\tabcolsep}{8pt}
\begin{tabular}{cccccc}
\hline
Systems & $d_f$ & $\nu$ & $\beta/\nu$ & $\gamma/\nu$ & $\tau$ \\
\hline
Poisson & $1.8955(25)$ & $1.3730(82)$ & $0.1055(24)$ & $1.8109(79)$ & $2.0383(4)$ \\
SHU $\chi = 0.20$ & $1.9090(33)$ & $1.3498(12)$ & $0.0938(38)$ &$1.7841(90)$ & $2.1320(8)$ \\
SHU $\chi = 0.40$ & $1.8971(9)$ & $1.3357(3)$ & $0.1044(13)$ &$1.7962(15)$ & $2.0708(3)$ \\
SHU $\chi = 0.49$ & $1.8959(5)$ & $1.3350(3)$ & $0.1061(7)$ &$1.7955(8)$ & $2.0567(2)$ \\
2D lattice perc. & $91/48\approx1.8958$ & $4/3\approx1.3333$ & $5/48\approx0.1042$ &$43/24\approx1.7917$ & $187/91\approx2.0549$ \\
\hline
\end{tabular}
\label{table:crit_exp}
\end{table*}

Among these exponents, $\nu$ is particularly informative for identifying possible changes in universality induced by long-range disorder. In particular,  the Weinrib--Halperin (WH) framework indicates that sufficiently long-ranged correlations can lead to a relevant perturbation to the standard percolation fixed point, with the correlation-length exponent $\nu$ directly modified by the correlated disorder \cite{Weinrib_correlated, WH_criterion}. 

As shown in Table~\ref{table:crit_exp}, the estimates of $\nu$ for the SHU networks with $\chi=0.40$ and $0.49$ are consistent with the standard 2D percolation value $\nu=4/3$. Likewise, the estimates of $d_f$, $\beta/\nu$, and $\gamma/\nu$ remain close to the corresponding standard values. The $\tau$ estimate for the SHU network with $\chi=0.40$ is slightly larger than the standard value. However, $\tau$ is more sensitive to the choice of cluster-size cutoffs and system size, which may account for this deviation. Overall, these results provide strong evidence that the high-$\chi$ SHU networks belong to the standard 2D percolation universality class.

The situation is less clear for the Poisson and SHU $\chi=0.20$ systems. The estimates of $\nu$ exhibit a systematic upward trend relative to the standard value $\nu=4/3$ as the suppression of long-wavelength fluctuations weakens, while $d_f$ and the exponent ratios remain comparatively close to their standard 2D percolation values. There is relevant precedent for such a pattern in studies of correlated percolation. Previous numerical studies of long-range correlated percolation have reported regimes in which $\nu$ departs from its uncorrelated value while $d_f$ and exponent ratios such as $\beta/\nu$ and $\gamma/\nu$ remain approximately unchanged, with systematic changes in these quantities appearing only for stronger correlations \cite{Prakash,Schrenk_long_range}.
Therefore, such a pattern can in principle occur when the disorder becomes relevant but is not yet sufficiently long ranged to induce comparable shifts in the exponent ratios and $d_f$. However, corrections to scaling may also produce apparent shifts in the effective critical exponents at finite system sizes \cite{Cardy_1996, Ziff_correction}. To distinguish between these two possibilities, we perform a scaling analysis based on the WH argument \cite{WH_criterion,Weinrib_correlated}, as described next.

\subsection{Disorder relevance and universality class}\label{subsec: scaling and universality}
Weinrib and Halperin's extension of the Harris criterion shows that disorder is irrelevant if fluctuations of a coarse-grained local control parameter decay faster than the distance from criticality \cite{WH_criterion,Weinrib_correlated}. For percolation, this can be formulated in terms of fluctuations of the block-averaged occupation variable. In our model, the quenched spatial disorder enters through the bond-dependent occupation probabilities $p_{ij}(z)$. We therefore use $p_{ij}(z_c)$ as a local proxy for the disorder controlling the percolation transition and examine how its fluctuations decay under spatial coarse-graining.

\begin{figure}[h]
    \centering
    \includegraphics[width=\columnwidth]{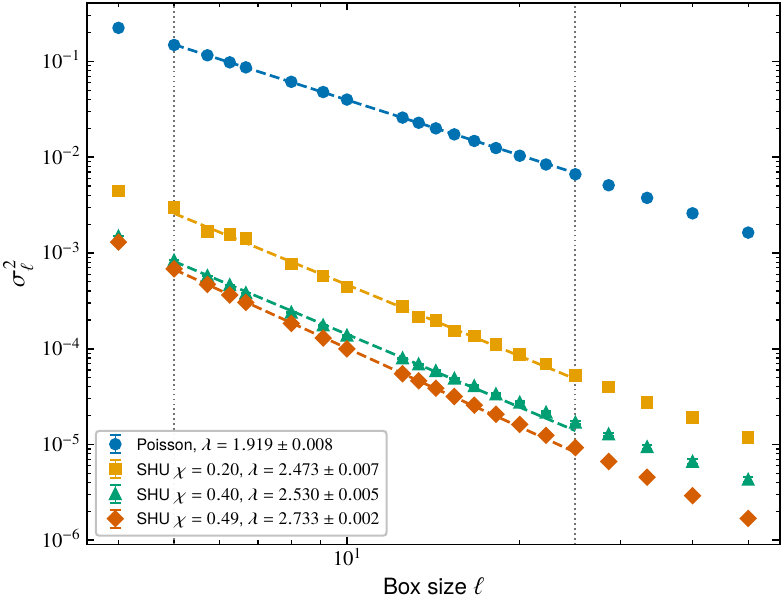}
    \caption{Block variance $\sigma_{\ell}^2$ of the coarse-grained bond probability field $p_{ij}(z_c)$ as a function of block size $\ell$ for the Poisson and SHU networks with different $\chi$. Dashed lines represent power-law fits $\sigma_{\ell}^2\sim \ell^{-\lambda}$ over the range $5\leq\ell\leq25$. Results are averaged over 20 disordered realizations with $L=200$. }
    \label{fig: res L_200}
\end{figure}

\begin{figure}[h]
    \centering
    \includegraphics[width=\columnwidth]{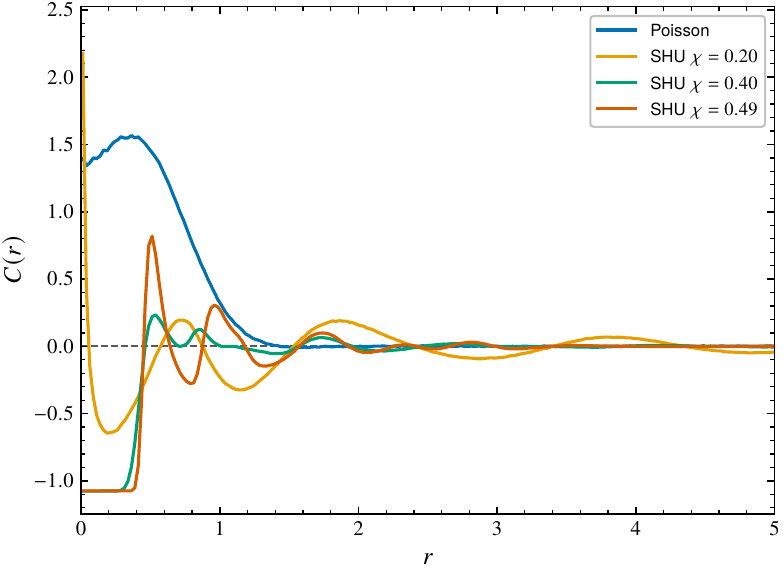}
    \caption{Connected correlation function $C(r)$ of the weighted bond field $p_{ij}(z_c)$ for the Poisson and SHU networks with different $\chi$. SHU networks exhibit oscillatory correlations, and therefore, the scaling of $\sigma_\ell^2$ cannot be characterized by a single decay exponent of $C(r)$. Results are averaged over 20 disorder realizations for systems with $L=200$.}
    \label{fig:Cr}
\end{figure}

We define the coarse-grained disorder field within a block $B$ as
\begin{equation}
    \bar{p}_{B}(z_c) = \frac{1}{\ell^2} \sum_{e_{ij}\in B} p_{ij}(z_c)
\end{equation}
where $B$ is chosen to be a square window of side length $\ell$ for convenience, and the position of each bond is defined by its midpoint. We calculate the variance of $\bar{p}_{B}(z_c)$, denoted by $\sigma_\ell^2$, as a function of $\ell$ and average over disorder realizations. Figure~\ref{fig: res L_200} shows the resulting scaling of $\sigma_\ell^2$ for all four systems. We restrict the fits to an intermediate range of $\ell$ to exclude small blocks containing only a few edges and large blocks strongly affected by the finite system size. The resulting block variance is well described by a power law,
\begin{equation}
    \sigma_\ell^2 \sim \ell^{-\lambda}
\end{equation}
with decay exponent $\lambda$. For the standard long-range correlated percolation problem, where the disorder correlation function decays as a power law, the asymptotic scaling of the block variance follows directly from the correlation function \cite{Weinrib_correlated}. In the SHU networks considered here, however, the correlation function of the weighted bond field $p_{ij}(z_c)$ is oscillatory. Figure~\ref{fig:Cr} shows the connected correlation function $C(r)$ of the weighted bond field $p_{ij}(z_c)$ for all systems. Since the block variance is given by a window-weighted integral of $C(r)$ \cite{Weinrib_correlated}, the alternating positive and negative correlations lead to cancellations. Consequently, the scaling of $\sigma_\ell^2$ cannot be characterized by a single decay exponent of $C(r)$, as in standard correlated percolation where $C(r)\sim r^{-a}$. Instead, it must be determined numerically, as we did here.

For standard 2D percolation, the correlation length scales as $\xi\sim |t|^{-\nu_0}$, or equivalently
$|t|\sim\xi^{-1/\nu_0}$, where $t=z-z_c$ denotes the distance from criticality. On the scale of the correlation length, the fluctuations of the coarse-grained disorder field scale as
\begin{equation}
    \sigma_\xi \sim \xi^{-\lambda/2}.
\end{equation}
Following the WH argument, disorder is irrelevant if
these fluctuations decay faster than the distance from criticality, such that
\begin{equation}
    \frac{\sigma_\xi}{|t|}
    \sim
    \xi^{1/\nu_0-\lambda/2}
\end{equation}
vanishes as $\xi\rightarrow\infty$ \cite{Weinrib_correlated}.
Within this scaling argument, the irrelevance condition is therefore
\begin{equation}
    \lambda>\frac{2}{\nu_0}.
\end{equation}
For standard 2D percolation, $\nu_0=4/3$, giving the criterion
$\lambda>3/2$. We summarize the estimated $\lambda$ of all systems for $L=100$ and $200$ in Table~\ref{tab: res}.

As a benchmark, we also include Bernoulli percolation of Delaunay triangulation networks built from Poisson point configurations, since this transition is known to belong to the standard 2D universality class \cite{Bernoulli_Poisson}. For Bernoulli percolation, $p_{ij}(z_c)=p_c$ for all edges, so $\bar{p}_{B}=p_c E/\ell^2$ where $E$ is the number of edges in the block. Imposing periodic boundary conditions on each block gives $E=3V$ where $V$ is the number of vertices (points). Since $\text{Var}(V)\propto\ell^2$ for Poisson configurations at large $\ell$, 
\begin{equation}
    \sigma^2_\ell=\frac{9p_c^2}{\ell^4}\operatorname{Var}(V)\sim\ell^{-2}
\end{equation}
giving $\lambda=2$. The numerical estimates in Table~\ref{tab: res} are consistent with this analytical value; the small deviations are due to finite-size and boundary effects. 

\begin{table}[h]
    \centering
    \caption{Summary of $\lambda$ for all cases.}
    \begin{tabular}{ccc}
        \hline
        Systems & $L=200$ & $L=100$ \\ \hline
        Poisson (Bernoulli) & $1.902(6)$ & $1.918(17)$ \\ \hline
        Poisson & $1.919(8)$ & $1.927(17)$ \\ \hline
        SHU $\chi=0.20$ & $2.473(7)$ & $2.593(13)$ \\ \hline
        SHU $\chi=0.40$ & $2.530(5)$ & $2.585(12)$ \\ \hline
        SHU $\chi=0.49$ & $2.733(2)$ & $2.766(7)$ \\ \hline
    \end{tabular}
    \label{tab: res}
\end{table}

As shown in Table~\ref{tab: res}, $\lambda$ generally increases with the degree of hyperuniformity of the underlying point configuration. Most importantly, all systems exhibit $\lambda>3/2$, and this qualitative conclusion remains unchanged whether $L=100$ or $L=200$. For the Poisson and SHU $\chi=0.20$ networks of particular interest here, the estimated $\lambda$ values are comparable to or larger than that of the Bernoulli benchmark, which belongs to the standard 2D percolation universality class. Within this WH-based analysis, these results are therefore consistent with the disorder being irrelevant and support the interpretation that the Poisson and SHU $\chi=0.20$ networks also belong to the standard 2D percolation universality class.

\begin{figure}[h]
\centering
\subfloat[Poisson networks: $\nu$]{
\includegraphics[width=0.46\textwidth]{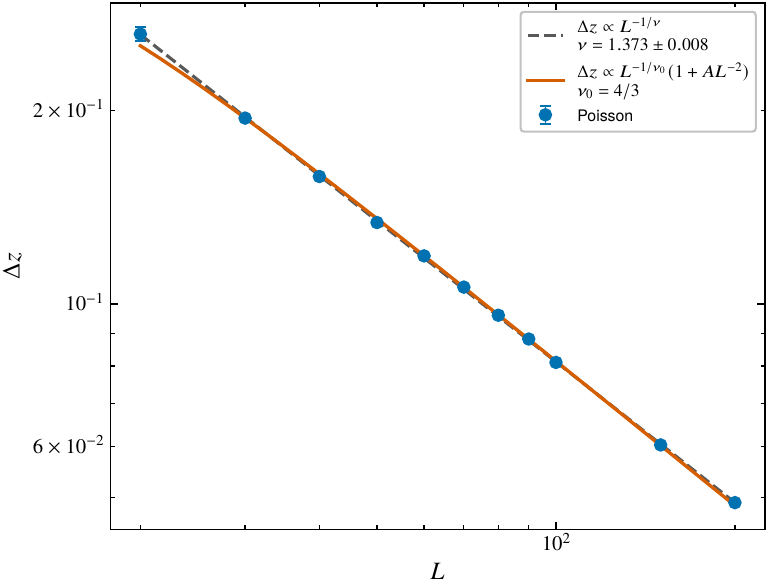}
}
\hfill
\subfloat[SHU networks with $\chi=0.20$: $d_f$]{
\includegraphics[width=0.46\textwidth]{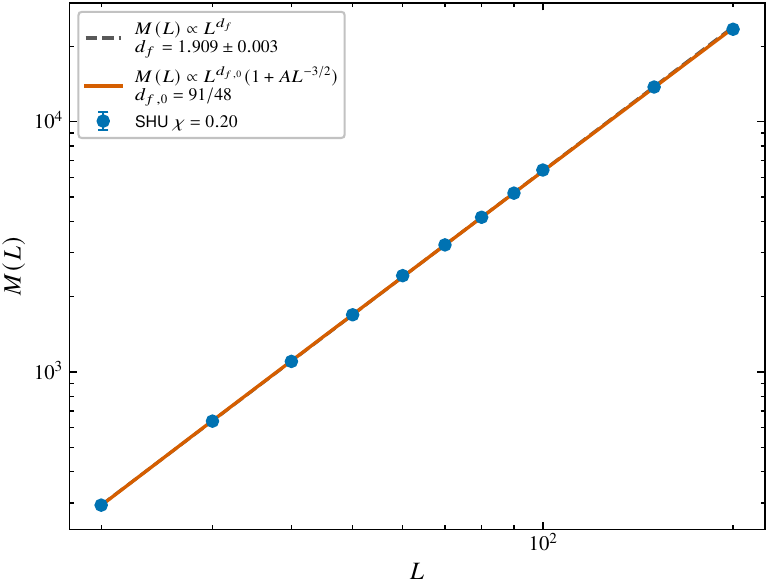}
}
\caption{Corrections to scaling of (a) the exponent $\nu$ of Poisson networks and (b) the fractal dimension $d_f$ of the SHU networks with $\chi=0.20$. The simulation results are consistent with leading-order corrections to scaling using the standard 2D percolation exponents $\nu_0=4/3$ and $d_{f,0}=91/48$.}
\label{fig: scaling corrections}
\end{figure}

Indeed, we find that the discrepancies observed in Table~\ref{table:crit_exp} are consistent with finite-size scaling corrections. For 2D percolation with periodic boundary conditions, the wrapping probability exhibits a leading correction \cite{Hovi_correction, Feng_2008_correction, NZ_alg}
\begin{equation}
    R_{L}(x)=R(x)+L^{-2}G(x)+\cdots
\end{equation}
where $R(x)$ is the asymptotic scaling function, $G(x)$ is the scaling function associated with the leading correction term, and $x=\left(z-z_c\right)L^{\frac{1}{\nu}}$. Since the width of the transition $\Delta z$ is defined by two fixed wrapping probabilities $R_1$ and $R_2$, expansion to leading order gives
\begin{equation}
    \Delta z \propto L^{-\frac{1}{\nu}}\left(1+AL^{-2}+\cdots\right)
\end{equation}
where $A$ is a non-universal constant. In addition, for observables related to the cluster-size distribution $n_s$, the correction assumes the form \cite{St94, Ziff_correction}
\begin{equation}
    n_s(z_c)\sim s^{-\tau}\left(1+Bs^{-\Omega}+\cdots\right)
\end{equation}
where $\Omega$ is the correction exponent. Since $\Omega=\omega/d_f$, this corresponds to a correction term proportional to $L^{-\omega}$ with $\omega=3/2$ in 2D. Figure \ref{fig: scaling corrections} shows corrections to scaling for representative critical exponents of Poisson and SHU networks with $\chi=0.20$. Upon including corrections to scaling, the simulation data are consistent with the standard 2D percolation exponents. We verified that the remaining exponent estimates are also consistent with the corresponding 2D percolation values. We note that, although corrected scaling forms provide a useful consistency check, simultaneous fitting of the critical exponent and correction parameters leads to strong parameter correlations over the accessible range of system sizes. We therefore use the leading-order scaling analysis to estimate the critical exponents, while the corrected forms are used to test consistency with the asymptotic 2D percolation values.

\begin{figure}[h!]
\centering
\subfloat[Scaling relation: $d_f=d-\beta/\nu$]{
\includegraphics[width=0.46\textwidth]{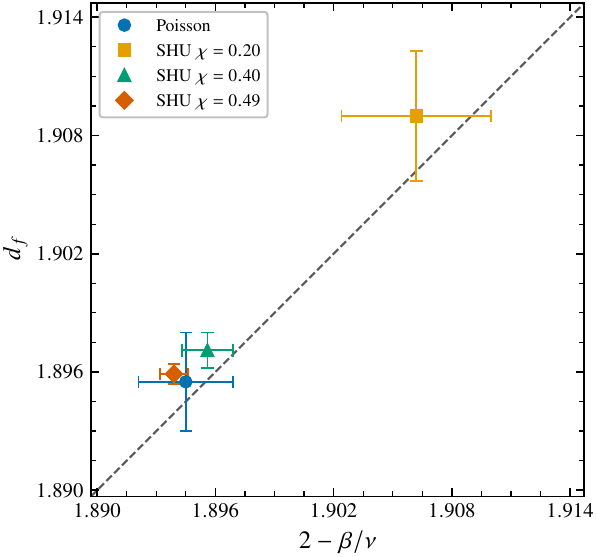}
}
\hfill
\subfloat[Scaling relation: $\gamma/\nu = 2d_f -2$]{
\includegraphics[width=0.46\textwidth]{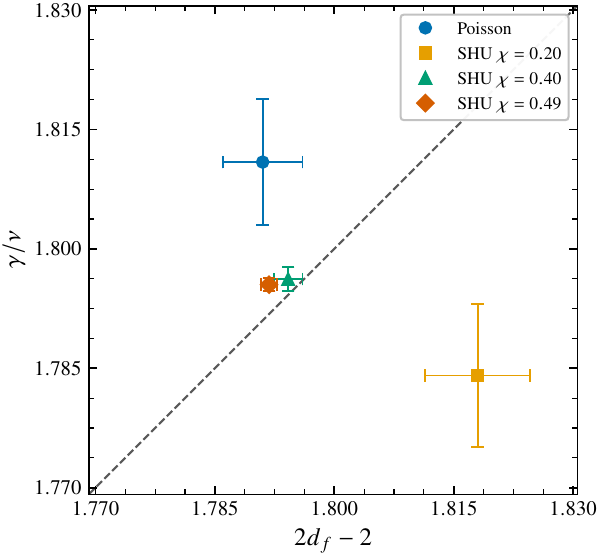}
}
\caption{Consistency checks of the scaling relations (a) $d_f=d-\beta/\nu$ and (b) $\gamma/\nu=2d_f-d$, where $d=2$, for all networks. The dashed lines indicate exact agreement between the independently determined critical exponents and the corresponding scaling relations.}
\label{fig: scaling relations}
\end{figure}

Finally, we examine the scaling relations among the independently determined critical exponents in Table~\ref{table:crit_exp}. Figure~\ref{fig: scaling relations} presents consistency checks for the scaling relations \cite{St94}
\begin{equation}
    d_f=d-\frac{\beta}{\nu}, \qquad
    \frac{\gamma}{\nu} = 2d_f - d
\end{equation}
where $d=2$ is the spatial dimension. The relation $d_f=d-\beta/\nu$ is closely satisfied for all systems. The relation $\gamma/\nu=2d_f-d$ is also well satisfied for the high-$\chi$ SHU networks, whereas larger deviations remain for the Poisson and SHU $\chi=0.20$ networks, consistent with the stronger finite-size corrections discussed above.

\section{Conclusion \& Discussion}\label{sec:conclusion_and_discussion}
In this work, we examine the percolation properties of Delaunay triangulation networks derived from Poisson and SHU point configurations with various $\chi$ under a distance-dependent bond-percolation model [Eq.~(\ref{eqn:perco_model})] that captures density information crucial to hyperuniformity. We also extend the classical Newman-Ziff algorithm to non-Bernoulli percolation problems.

Our results show that the percolation threshold $z_c$ depends strongly on the degree of hyperuniformity of the underlying point configurations, and increasing the short-range order gives rise to a lower percolation threshold. The lower percolation thresholds observed in stealthy systems are also closely linked to their ``bounded-hole'' property \cite{Zh17}. Furthermore, we find that all systems considered are consistent with the standard 2D percolation universality class, although the Poisson and low-$\chi$ SHU networks exhibit stronger finite-size corrections to scaling.

This work uses Eq.~(\ref{eqn:perco_model}) to incorporate geometric information into the percolation model by mapping the Delaunay edge lengths directly onto bond occupation probabilities. The linear kernel provides a simple and transparent choice in which shorter edges are preferentially occupied. Nonetheless, other physically well-motivated monotonic distance-dependent kernels, such as exponential forms, could be considered depending on the application. We expect the qualitative hierarchy of percolation thresholds to persist for a broad class of kernels that preferentially occupy shorter edges, since this trend originates from the suppression of large voids and long bridging distances in the more ordered SHU configurations. However, the numerical values of the percolation threshold and other nonuniversal quantities, such as finite-size amplitudes, will generally depend on the specific choice of kernel. By contrast, whether the universality class changes cannot be inferred from the microscopic form of the kernel alone. For any specified distance-dependent kernel, the corresponding bond-disorder field can be defined and the WH-based scaling analysis introduced in Sec.~\ref{subsec: scaling and universality} can be applied to determine whether the induced disorder is relevant to the standard 2D percolation fixed point.

We mainly focus on networks constructed using the Delaunay triangulation, which provides a parameter-free representation that captures local geometry while retaining a planar embedding. Nonetheless, a graph representation inevitably discards certain aspects of the underlying spatial structure \cite{HU_Eli_2, HU_network}, and alternative network constructions may induce different effective disorder correlations. The planarity of the Delaunay triangulation also enables straightforward detection of wrapping and finite-size-scaling analysis, whereas non-planar spatial networks may require alternative approaches, such as the generating-function formalism \cite{Newman_RG, Newman_perco}. Investigating how network construction affects the resulting percolation behavior therefore remains an interesting direction for future work.

Furthermore, our findings open up the possibility of optimizing the resilience of disordered planar networks that are statistically homogeneous and isotropic. Percolation has been widely used to model network resilience, including random breakdowns of the Internet \cite{Internet_perco, Newman_perco}. A network is considered resilient if the removal of a large fraction of vertices or edges does not lead to fragmentation of the spanning cluster. In our model, edge removal is distance-dependent, and a more resilient network corresponds to one that maintains spanning connectivity down to a lower coupling scale. Hence, Delaunay networks derived from hyperuniform point configurations are more resistant to distance-dependent edge removal than those derived from uncorrelated point configurations.

In future work, we will investigate the percolation behavior of other disordered systems with ``bounded holes,'' such as random sequential addition (RSA) packings \cite{torquato2006random, zhang2013precise}. We hypothesize that these systems will exhibit lower percolation thresholds than the corresponding Poisson system, owing to their enhanced short-range order and suppressed large void fluctuations. 
\\

\begin{acknowledgments}
This work was supported by the Army Research Office under Cooperative Agreement Number W911NF-22-2-0103.
\end{acknowledgments}

\bibliography{reference2}

\end{document}